%% file: proceedings.tex
\documentclass{PoS}

\usepackage{amsmath}
\usepackage{amsthm}
\usepackage{algorithm}
\usepackage{algorithmic}

\theoremstyle{definition}

\let\phi=\varphi

\numberwithin{equation}{section}

\numberwithin{algorithm}{section}

\title{Aggregation-based Multilevel Methods for\\ Lattice QCD}

\ShortTitle{Multilevel Methods}

\author{A.\ Frommer\\
       Department~of~Mathematics, 
       Bergische~Universit\"at~Wuppertal, 42097~Germany\\
       E-mail: \email{frommer@math.uni-wuppertal.de}}
       
\author{K.\ Kahl\\
       Department~of~Mathematics, 
       Bergische~Universit\"at~Wuppertal, 42097~Germany\\
       E-mail: \email{kkahl@math.uni-wuppertal.de}}
       
\author{S.\ Krieg\\
			 Department~of~Physics, Bergische~Universit\"at~Wuppertal, 42097~Germany\\
			 \& J\"ulich~Supercomputing~Centre, Forschungszentrum~J\"ulich, 52428~Germany\\
       E-mail: \email{s.krieg@fz-juelich.de}}
       
\author{B.\ Leder\\
       Department~of~Mathematics, 
       Bergische~Universit\"at~Wuppertal, 42097~Germany\\
       E-mail: \email{leder@math.uni-wuppertal.de}}

\author{\speaker{M.\ Rottmann}\\ 
        Department~of~Mathematics, 
	Bergische~Universit\"at~Wuppertal, 42097~Germany\\
        E-mail: \email{rottmann@math.uni-wuppertal.de}}

\abstract{
In Lattice QCD computations a substantial amount of work is spent in
solving the Dirac equation.
In the recent past it has been observed that conventional
Krylov solvers tend to critically slow down for large
lattices and small quark masses.
We present a Schwarz alternating procedure (SAP) multilevel method
as a solver for the Clover improved Wilson discretization of the Dirac equation.
This approach combines two components (SAP and algebraic multigrid)
that have separately been used in lattice QCD before.
In combination with a bootstrap setup procedure we show that considerable speed-up
over conventional Krylov subspace methods for realistic configurations can be achieved.
}

\FullConference{ The XXIX International Symposium on Lattice Field Theory - Lattice 2011\\
July 10-16, 2011\\
Squaw Valley, Lake Tahoe, California}

\begin{document}

\section{Introduction}
\vspace{-1.0em}
The most costly task in lattice QCD computations is the solution of large sparse linear systems of equations
\begin{equation}\label{eq:linearsystem}
D z = b,
\end{equation} where $D$ is a discretization of the Dirac operator.
Here we consider the Wilson discretization $D = D(U) + m$ which
couples only nearest neighbors and
depends on a gauge field $U$ and a mass parameter $m$. Usually $z$ is calculated by a Krylov subspace method (e.g.\ CGN, GCR, BiCGStab).
Those methods suffer from critical slowing down
when approaching the critical mass as well as lattice spacing $a=0$.
Thus it is of utmost importance to develop preconditioners for these methods that
remedy these scaling problems. 

In the recent past preconditioners based on domain decomposition (DD) for the solution of~\eqref{eq:linearsystem} 
have been proposed in~\cite{Luscher:2003qa}. Although DD methods excel in supercomputing environments due to 
their high inherent parallelism they are unable to remedy the scaling problems completely unless
they are combined with a multilevel approach. Thus we combine the DD approach with an algebraic multigrid hierarchy based on 
a bootstrap aggregation framework~\cite{KahlBootstrap, Brezina2005}. Our approach is similar in construction to
the one introduced in~\cite{Osborn:2010mb, Babich:2010qb, Brannick:2007ue} 
where it has been shown that using such algebraic multigrid approaches can remedy the scaling problems in QCD computations.
As in ~\cite{Osborn:2010mb, Babich:2010qb, Brannick:2007ue} we obtain a multilevel hierarchy
using non-smoothed aggregation. The difference is that we replace the multigrid smoother
by a DD approach, expecting a gain in efficiency on highly parallel machines.

In section~\ref{domain_decomposition} we introduce the concept of DD methods before we explain the construction of our algebraic multilevel
method in some detail in section~\ref{algebraic_mg}. Thereafter we give numerical results of our method in section~\ref{results} and 
finish with some concluding remarks.
\vspace{-1.5em}
\section{Domain Decomposition}
\label{domain_decomposition}
\vspace{-1.0em}
Domain decomposition methods were developed as iterative solvers for linear systems
arising from discretizations of PDEs.
The main idea consists of solving the
system on the whole domain by repeatedly solving smaller systems with less degrees of freedom on smaller subdomains.

\begin{figure}[b]
  \begin{minipage}[t]{0.45\textwidth}
    \centering \scalebox{0.85}{\input{img/redblack_small.tex}}
    \caption{block decomposed lattice (reduced to 2D) with 2 colors}
    \label{redblack_small}
  \end{minipage}
\hfill
  \begin{minipage}[t]{0.45\textwidth}
    \centering \scalebox{0.85}{\includegraphics[scale=0.6]{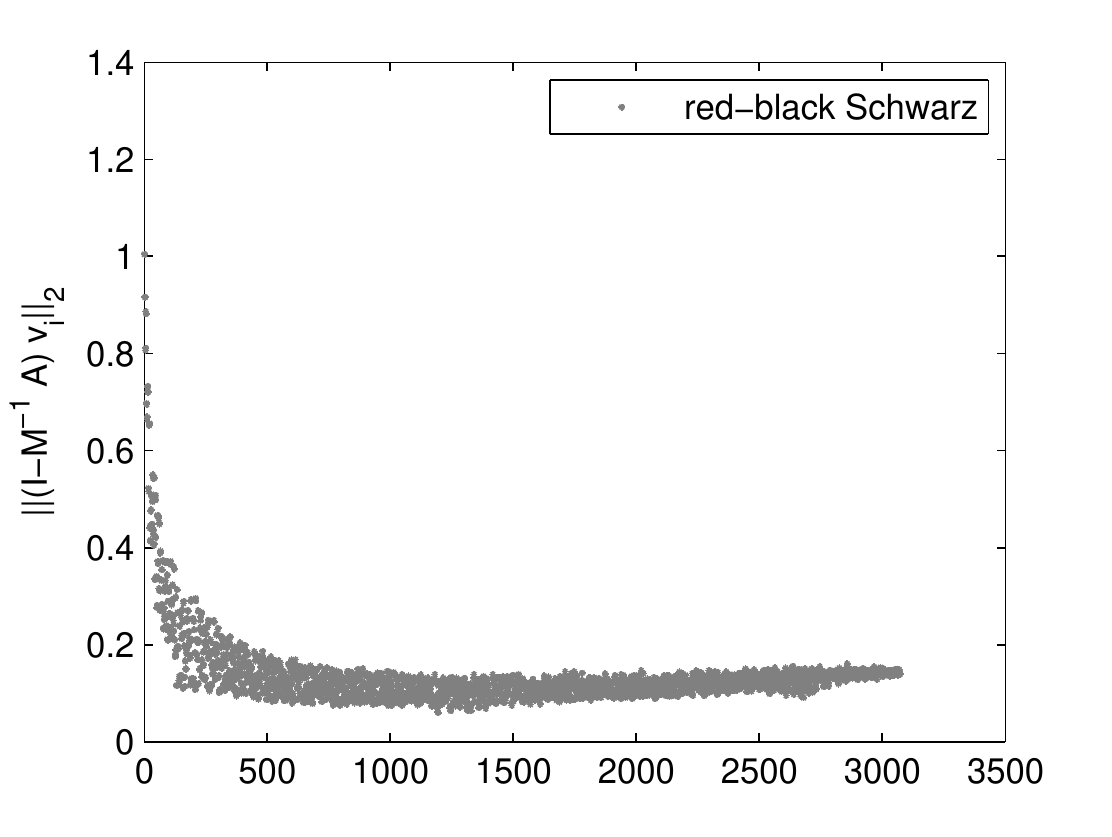}}
    \caption{error component reduction in terms of EVs on a $4^4$ lattice with $2^4$ blocks, EWs sorted by magnitude}
    \label{errSAP}
  \end{minipage}
\end{figure}

Consider a block decomposition $ \{ \mathcal{L}_i : i=1,\ldots,k \} $ of a lattice $ \mathcal{L} $
(Figure \ref{redblack_small} illustrates a 2D example).
The corresponding trivial embeddings and block solvers are denoted by $ I_{\mathcal{L}_i} : \mathcal{L}_i \rightarrow \mathcal{L} $
and $B_i = I_{\mathcal{L}_i} [ I_{\mathcal{L}_i}^T D I_{\mathcal{L}_i} ]^{-1} I_{\mathcal{L}_i}^T $.
Note that the trivial embedding $I_{\mathcal{L}_i}$ is just the restriction
of the identity on $\mathcal{L}$ to $\mathcal{L}_i$. Then one iteration of a domain decomposition method consists of 
solving each of the block systems $e \leftarrow B_i r $, interleaved with a number of residual updates $r = b - Dz$. In the two extreme cases
where one does only one residual update before solving all block systems or when the residual is updated after each 
block solution, the corresponding error propagators are given by
\begin{equation} \label{additive}
1 - (\sum_{i=1}^k B_i) A \quad \text{and} \quad \prod_{i=1}^k \left( 1 - B_i A \right).
\end{equation}
These methods go back to H.~Schwarz~\cite{Schwarz1870} and thus are
called additive Schwarz and multiplicative Schwarz method, respectively. The block systems of the additive variant 
can be solved simultaneously while the multiplicative variant is inherently sequential. Its advantage is that it spreads the information 
faster on the lattice as a solution of a block system uses previous solutions of other block systems.

The two methods can be combined to exploit advantages of both methods.
Coloring the blocks such that no adjacent blocks have the same color, a residual update on one block 
no longer influences the residual on blocks of the same color.
Thus it suffices to perform the update once for each color. All blocks of the same color
can then be computed simultaneously as described in Algorithm~\ref{rbsap}.
Such a DD approach has been applied to solve~\eqref{eq:linearsystem} in~\cite{Luscher:2003qa}, where it has been named
Schwarz Alternating Procedure (SAP).

\begin{algorithm}[t]
  \caption{Red-Black Schwarz}
  \label{rbsap}
  \begin{algorithmic}[1]
    \FOR{$c=1$ to $2$}
      \STATE $r \leftarrow b-Dz$
      \FORALL{$i \in \{1,...,k\}$ with $\mathit{color}(i) = c$}
	\STATE $z \leftarrow z+B_i r $
      \ENDFOR
    \ENDFOR
  \end{algorithmic}
\end{algorithm}

Typically the solution of the block system $e = B_i r$ is approximated by a few iterations of a Krylov subspace method (e.g.\ GMRES),
and the DD method itself is in turn used as a preconditioner for a (flexible) Krylov subspace method.
As illustrated in Figure~\ref{errSAP} we observe that SAP is able to reduce error components belonging to a large part of the spectrum very well
but a small part belonging to eigenvectors (EVs) to small eigenvalues (EWs) remains intractable.
For larger configurations the number of EWs with small magnitude of the Dirac operator gets larger, which yields an explanation
why SAP is not able to remedy the scaling problem as the number of intractable eigenvectors increases as well.
Though, the seen behavior of damping large EVs is desirable for an iterative method to be used as a smoother in a multigrid method
and motivated us to use it in this context.
\vspace{-1.5em}
\section{Algebraic Multigrid}
\label{algebraic_mg}
\vspace{-1.0em}
A multigrid method typically consists of a simple iterative method called smoother and complementary coarse grid correction.
As motivated in section~\ref{domain_decomposition} we deem SAP suitable for the use as a smoother since it is cheap to compute and
reduces the error efficiently on a large part of the spectrum. The main idea of multigrid is to treat the error that is left
after a few iterations of the smoother within a smaller subspace where the troublesome error components can be approximated.
More precisely we want to define and solve a ``coarse'' linear system
\begin{equation} \label{coarse_eq}
D_c z_c = b_c
\end{equation}
with a much smaller operator $D_c: \mathbb{C}^{n_c} \rightarrow \mathbb{C}^{n_c}$
in order to reduce the error on the critical part of~\eqref{eq:linearsystem}.
To this purpose we have to define linear maps $R: \mathbb{C}^{n} \rightarrow \mathbb{C}^{n_c}$ to restrict information 
based on the current residual $r = b-Dz$ to the subspace and a linear map $P: \mathbb{C}^{n_c} \rightarrow \mathbb{C}^{n}$ to interpolate
the information that we obtain from solving \eqref{coarse_eq} back to $\mathbb{C}^{n}$ where \eqref{eq:linearsystem} is given.
This yields a subspace correction
\begin{equation} \label{coarse_grid_correction}
z \leftarrow z + P D_c^{-1} R r
\end{equation}
with the corresponding error propagator $ 1 - P D_c^{-1} R D $. As $D_c$ should resemble the action of D on the troublesome 
subspace approximated by $\mathit{span}(P)$, the action of $D_c$ is chosen as
the action of $D$ on interpolated vectors which are restricted afterwards.
Formally this amounts to a Petrov-Galerkin formulation of the coarse operator as
$ D_c = RDP $. With this choice of $D_c$ the error propagator of the subspace correction is given by 
$ 1 - P (R DP)^{-1} R D $. In order to benefit from such a subspace correction, solving \eqref{coarse_eq} has to be much cheaper than
solving \eqref{eq:linearsystem}. That is $n_c$ should be small compared to $n$, and $D_c$ should be sparse.
As the dimension of the troublesome subspace grows with $n$ (cf. \cite{Luscher:2007se})
we do not want to fix $n_c$ (like in deflation methods)
but want to find a sparse description of $D_c$ on that subspace.

Once $D_c$ is found a basic two level algorithm consists of the alternating application
of smoother and subspace correction.
This procedure can be recursively extended by formulating a two level algorithm
of this kind for the computation of \eqref{coarse_eq}
until we get an operator which is small enough to solve \eqref{coarse_eq} directly.


\textbf{Aggregation Based Interpolation:} We decided to adjust an aggregation based interpolation,
that in turn yields the subspace correction, as the complementary component to the SAP smoother.
Due to the fact that the coarse grid correction in \eqref{coarse_grid_correction} only acts on error components in $\mathit{range}(P)$,
it should approximate the subspace spanned by eigenvectors to eigenvalues of small magnitude of $D$ (cf.\ Figure \ref{errSAP}).
In the multigrid literature, $P$ is built
from right, and $R$ from left EVs
corresponding to EWs of small magnitude of $D$.
Due to the spectral properties of the Wilson Dirac operator it is natural to choose
$ R = \gamma_5 P $. Furthermore, with additional assumptions on its structure we can choose $R = P^\dagger$ according to \cite{Babich:2010qb}.
Therefore we define the aggregates in such a way 
that there exists $ \gamma_5^c $ 
such that 
$\gamma_5 P = P \gamma_5^c$. Note, that with these assumptions on $P$ the error propagator 
of the subspace correction \eqref{coarse_grid_correction} is given by $z \leftarrow z + P (P^\dagger DP)^{-1} P^\dagger r$.

With this structure of $P$ in mind, we define its entries 
based on a set of test vectors $\{ v_1,\ldots, v_N \}$
whose span approximates the troublesome subspace
and a set of aggregates $ \{ \mathcal{A}_1,\ldots, \mathcal{A}_s \}$.
The aggregates 
can be realized as another block decomposition of the lattice. 
Note that the DD smoother and the interpolation do not have to share a common block decomposition.
\begin{figure}[t]
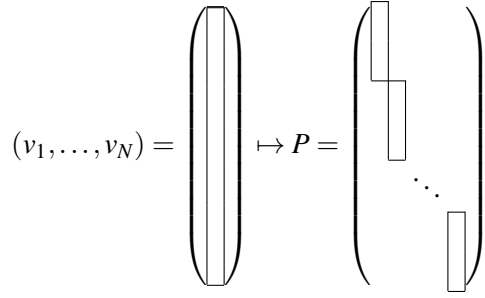


  \[ { (v_1,\ldots,v_N) = 
  \left( \begin{array}{|c|}
  \hline
  \cline{1-1}
  \multicolumn{1}{|c|}{ \; } \\
  \multicolumn{1}{|c|}{ \; } \\
  \multicolumn{1}{|c|}{ \; } \\
  \multicolumn{1}{|c|}{ \; } \\
  \multicolumn{1}{|c|}{ \; } \\
  \multicolumn{1}{|c|}{ \; } \\
  \multicolumn{1}{|c|}{ \; } \\
  \hline
  \end{array} \right)
  \mapsto
  P = \left( \begin{array}{c c c c c c c c}
    \cline{1-1}
    \multicolumn{1}{|c|}{ \; } & & & \\
    \multicolumn{1}{|c|}{ \; } & & & \\
    \cline{1-2}
    & \multicolumn{1}{|c|}{ \; } & & \\
    & \multicolumn{1}{|c|}{ \; } & & \\
    \cline{2-2}
    & & \ddots & \\
    \cline{4-4}
    & & & \multicolumn{1}{|c|}{ \; } \\
    & & & \multicolumn{1}{|c|}{ \; } \\
    \cline{4-4}
    \end{array} \right)
  }
  \]
  \caption{construction of the interpolation (operator based point of view)}
  \label{aggregation_matrix}
\end{figure} The interpolation $P$ is then given by decomposing the test vectors over the aggregates (cf.\ Figure \ref{aggregation_matrix}).
Hence $P$ is a linear map from the coarse grid to the fine grid, defined by
$$ Pe_j := I^T_{\mathcal{A}_{\lceil \frac{j}{N} \rceil} } v_{((j-1)\text{ mod }N)+1} \quad \text{for} \quad j = 1,\ldots,N \cdot s $$
where $e_j$ is the $j$-th unit vector. In order to have $ P^\dagger P = I $,
the test vectors are locally orthonormalized over the aggregates.
Note that due to the aggregation structure of $P$ and $R$ the sparsity/connection structure
of $D_c$ resembles the one of $D$, i.e., the corresponding graphs
of the operators are regular four dimensional grids.
Thus we can apply \eqref{coarse_grid_correction} recursively to \eqref{coarse_eq} and
obtain a hierarchy of coarser grids and coarser operators.
This construction of $P$ is similar to constructions found in
\cite{Osborn:2010mb, Babich:2010qb, Brannick:2007ue, Luscher:2007se}.
More precisely, the structure of the interpolation operators is identical
but the test vectors $v_i$ used to build them and thus the actions of the operators
are different.

%
%
%


\textbf{Bootstrap Setup:} A critical part of the construction of an efficient multigrid hierarchy
is the computation of the test vectors used in the definition of $P$.
The setup procedure we employ for this task is divided into two parts.
We start with a set of random test vectors $v_1,\ldots,v_N$
and apply a small number of smoother iterations $S^\nu$
to them. During this procedure the smoothed test vectors $S^\nu v_i$ are
kept orthonormal. After that a temporary operator
on the next coarser grid is computed according to the aggregation based construction.
This procedure is continued recursively on the coarser grids 
until we get an operator on the coarsest grid. 
Herein we restrict smoothed test vectors with $P^\dagger$
to the next coarser grid.
In what follows we omit the level indices for the sake of simplicity.

Since the subspace $\mathit{range}(P)$ should contain a significant part
of the eigenvectors to eigenvalues of small magnitude 
of the fine grid space $\mathbb{C}^n$, the EWs of
small magnitude of $D_c = P^\dagger D P$ should approximate those of $D$.
For the corresponding EVs the relations
$ P^\dagger P = 1 $ and $ P^\dagger(D P \phi - \lambda P \phi) = 0 $
imply
$ D P \phi \approx \lambda P \phi$.
Thus the second part of our setup procedure
starts with the calculation of $N$ approximations to the smallest EVs and EWs $ \{ ( \lambda_i, \varphi_i )\}$ of $P^\dagger D P$ 
by means of harmonic Ritz vectors and values.
The approximate EVs $\varphi_i$ for $i=1,\ldots,N$
are successively interpolated to the next finer grid and
smoothed towards their harmonic Ritz values
with some steps of SAP with iteration matrix $ S(\lambda_i)^\iota $. In this case the local inverses for the pair $ ( \lambda_i, \varphi_i ) $ are given by
$$ B_j(\lambda_i) = I_{\mathcal{L}_j} [ I^T_{\mathcal{L}_j} ( D - \lambda_i ) I_{\mathcal{L}_j} ]^{-1} I^T_{\mathcal{L}_j}. $$
The resulting set of vectors and the old test vectors
$ V := \{ S(\lambda_i)^\iota P \varphi_i : i=1,\ldots,N \} \cup \{ S^\nu v_j : j=1,\ldots,N \} $
are again reduced to $N$ vectors. In order to preserve the most significant information, the $N$ smallest singular values and their corresponding vectors
of the $2N \times 2N$ matrix
$ (DV)^\dagger DV $
are calculated and the corresponding $N$ linear combinations 
of the vectors of $V$ are the final vectors $\hat{V}$.
They define the interpolation from the next coarser grid to the current grid
and the operator on the next coarser grid. 
The second part of the setup executes this procedure exactly once,
starting on the coarsest grid.

\begin{figure}[t]
  \begin{minipage}[hbt]{0.46\textwidth}
    \centering \includegraphics[scale=0.50]{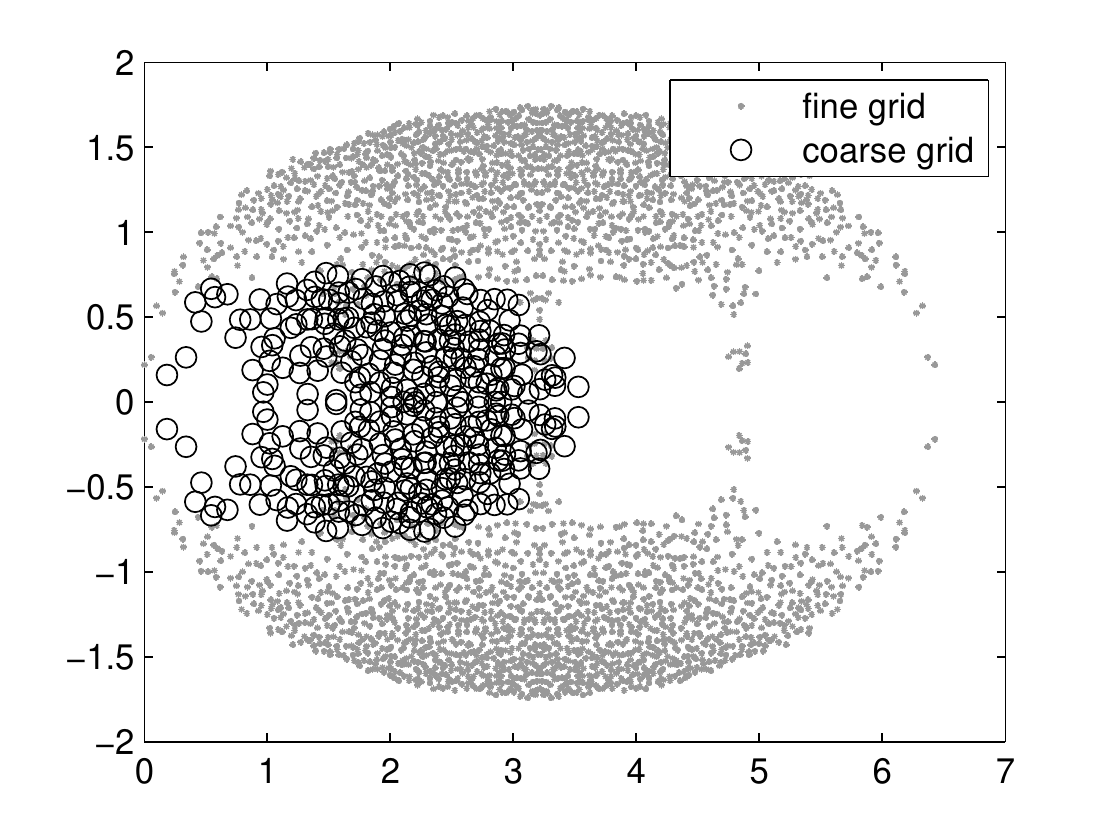}
    \caption{Spectra after the first setup step, $4^4$ Wilson, $2^4$ aggregate size}
    \label{imp0}
  \end{minipage}
\hfill
  \begin{minipage}[hbt]{0.46\textwidth}
    \centering \includegraphics[scale=0.50]{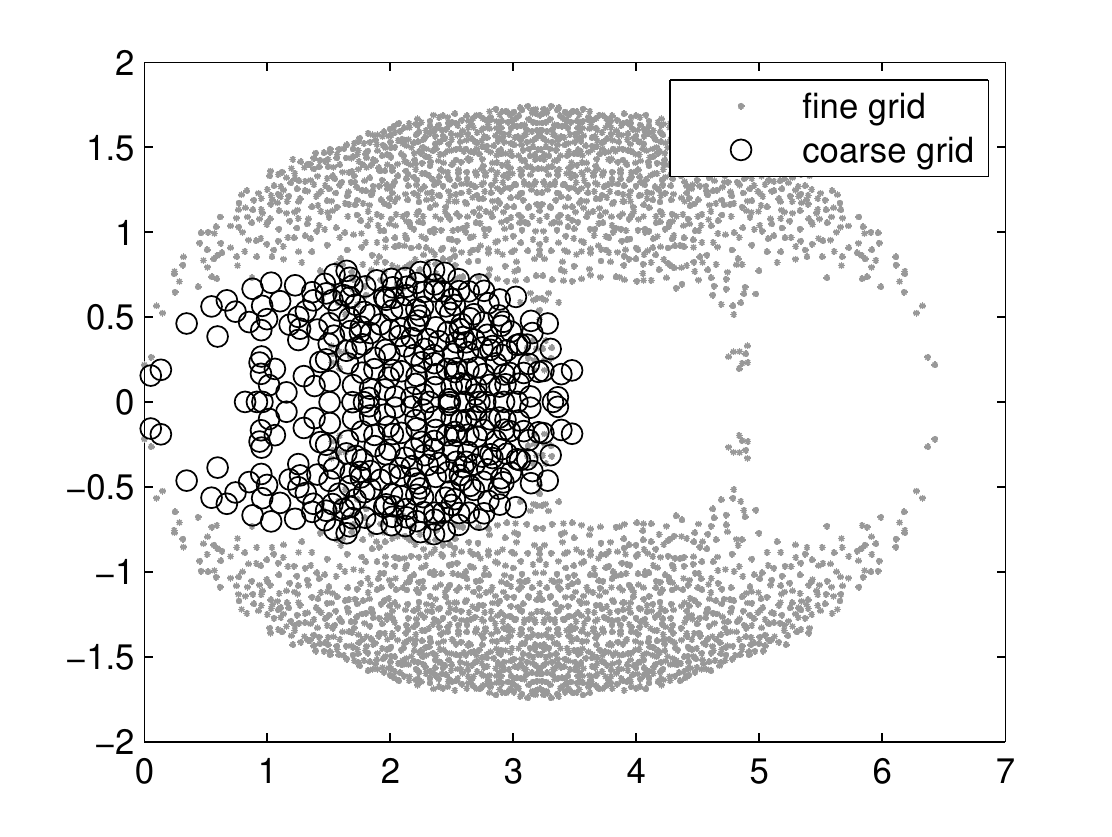}
    \caption{Spectra after the second setup step, $4^4$ Wilson, $2^4$ aggregate size}
    \label{imp1}
  \end{minipage}
\end{figure}

Figures \ref{imp0} and \ref{imp1} illustrate the 
influence of the second part of the setup procedure on
the coarse grid operator in a two level hierarchy.
The smallest eigenvalues of the fine grid operator 
are much better represented on the coarse grid.
For other setup procedures, see e.g.\ \cite{Osborn:2010mb, KahlBootstrap, Brezina2005}.
\vspace{-1.5em}
\section{Results}
\label{results}
\vspace{-1.0em}
Our adaptive DD multilevel solver ($\alpha$MG-DD) has been implemented in \texttt{C} using the parallelization interface of \texttt{MPI}.
Krylov subspace methods have been implemented within a common framework for a fair comparison.
For the numerical experiments we have combined our multigrid solver against CGN, i.e.\ CG on the normal equation.
All results were produced with a two level method.
The stopping criterion was to reduce the initial residual norm by a factor of at least $10^{10}$.
The solutions of the block systems within SAP were approximated by $3$ iterations of GMRES and
the coarse grid equation \eqref{coarse_eq} was approximately solved with $12$ iterations of GMRES.
For the outer flexible GMRES routine we chose a
restart length of $25$. All results have been computed on Juropa at the J\"ulich Supercomputing Centre.

\medskip
\begin{minipage}[t]{0.45\textwidth}
\leftskip=-2em{
  \begin{tabular}{c|c||c|c}
	  &            & FGMRES              &        \\
          &            & + $\alpha{}$MG-DD  & CGN    \\ \hline

    setup & timings    & 45.22s    & -      \\ \hline

    solve & iterations & 9       & 4476 \\
          & timings    & 6.97s   & 121.82s \\ \hline
	  
    total & timings  & 52.19s    & 121.82s
  \end{tabular}
\\

\small{\textbf{Table\;1:} $48 \times 24^3$ Clover Wilson-Dirac,
$\beta = 5.3$ $(1/a = 2.56 \text{ GeV})$, $\kappa = 0.13590$ ($m_\pi=630 \text{ MeV}$),
$c_{\text{sw}} = 1.90952$,
generated using public code with parameters from L.~Del~Debbio~\cite{DelDebbio:2006cn},
block-size $3^4$, coarsening $3^4 \times 6 \rightarrow 20 $,
$512$ cores (Juropa at JSC).}
}
\end{minipage} \hfill
\begin{minipage}[t]{0.45\textwidth}
  \begin{tabular}{c|c||c|c}
	  &            & FGMRES &         \\
          &            & + $\alpha{}$MG-DD & CGN    \\ \hline

    setup & timings    & 24.85s    & -      \\ \hline

    solve & iterations & 17     & 9181  \\
          & timings    & 5.05s   & 119.04s \\ \hline
	  
    total & timings  & 29.90s    & 119.04s
  \end{tabular}
\\

\small{\textbf{Table\;2:} $64 \times 32^3$ Wilson-Dirac
2HEX smeared tree level improved Clover,
$\beta = 3.5 \; (1/a = 2.130 \text{ GeV}) $,
$\kappa = 0.12646$ ($m_\pi=300 \text{ MeV} $),
$c_{\text{sw}} = 1.0$,
provided by the BMW~collaboration~\cite{Durr:2010vn, Durr:2010aw},
block-size $2^4$, coarsening $4^4 \times 6 \rightarrow 20 $,
$4096$ cores (Juropa at JSC).}
\end{minipage}

\medskip

In the cases shown our multigrid solver was $17$ and $24$ times faster
than CGN. Including the setup time we are still twice and four times as fast as CGN.
Since the setup has to be done only once the benefits of our approach are larger 
the more right hand sides there are to be solved.
\vspace{-1.5em}
\section{Summary and Outlook}
\label{summary}
\vspace{-1.0em}
The developed method combining DD techniques and algebraic multigrid shows great potential
to speed-up calculations of propagators in lattice QCD. Even for single right hand 
sides our method outperforms conventional Krylov subspace methods with the potential of an even
more significant speed-up when solving for many right hand sides. This result is mainly due to the
introduction of the highly parallel DD smoother and the bootstrap setup into the algebraic multigrid method. 
While the first speeds up the setup of the method \textit{and} the subsequent solution the latter significantly speeds 
up the setup of the multigrid hierarchy, which in general is a bottleneck in algebraic multigrid methods
especially compared to setup-free Krylov subspace methods. We are currently working on an optimized version of
the code that should be able to run on large-scale parallel machines and extend our testing of the method
towards larger lattices and lighter quark masses. In the near future we plan to incorporate our algorithm
into the production codes of our collaborators within SFB TR55.


\textbf{Acknowledgments:} This work is funded by Deutsche Forschungsgemeinschaft (DFG)
Transregional Collaborative Research Centre 55 (SFB TR55).




\bibliographystyle{utphys}
\bibliography{myrefs}


\end{document}

%% file: img/redblack_small.tex
\begin{picture}(0,0)%
\includegraphics{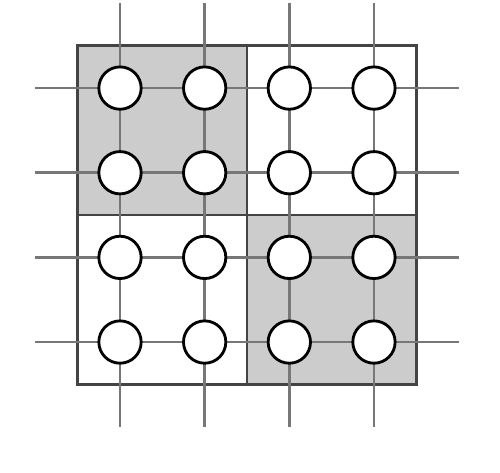}%
\end{picture}%
\setlength{\unitlength}{1782sp}%
\begingroup\makeatletter\ifx\SetFigFont\undefined%
\gdef\SetFigFont#1#2#3#4#5{%
  \reset@font\fontsize{#1}{#2pt}%
  \fontfamily{#3}\fontseries{#4}\fontshape{#5}%
  \selectfont}%
\fi\endgroup%
\begin{picture}(5295,5016)(526,-4819)
\put(541,-3031){\makebox(0,0)[b]{\smash{{\SetFigFont{9}{10.8}{\familydefault}{\mddefault}{\updefault}$\mathcal{L}_3$}}}}
\put(541,-1231){\makebox(0,0)[b]{\smash{{\SetFigFont{9}{10.8}{\familydefault}{\mddefault}{\updefault}$\mathcal{L}_1$}}}}
\put(5806,-1231){\makebox(0,0)[b]{\smash{{\SetFigFont{9}{10.8}{\familydefault}{\mddefault}{\updefault}$\mathcal{L}_2$}}}}
\put(5806,-3076){\makebox(0,0)[b]{\smash{{\SetFigFont{9}{10.8}{\familydefault}{\mddefault}{\updefault}$\mathcal{L}_4$}}}}
\put(3106,-4696){\makebox(0,0)[b]{\smash{{\SetFigFont{9}{10.8}{\familydefault}{\mddefault}{\updefault}$\mathcal{L}$}}}}
\end{picture}%